\documentclass[prr,aps,showpacs,twocolumn,amsmath,amssymb,floatfix,superscriptaddress]{revtex4-2}

\usepackage{hyperref}
\usepackage{graphicx}
\usepackage{bm}
\usepackage{xcolor}
\usepackage[utf8]{inputenc}
\usepackage{amssymb}
\usepackage{bbm}
\usepackage[normalem]{ulem}

\newcommand{\beq}{\begin{equation}}
\newcommand{\eeq}{\end{equation}}
\newcommand{\beqa}{\begin{eqnarray}}
\newcommand{\eeqa}{\end{eqnarray}}

\newcommand{\nn}{\nonumber}

\newcommand{\editE}[1]{#1}

\begin{document}

\title{\editE{Solitonic Andreev spin qubits from Andreev states in Corbino Josephson junctions}}

\author{Pablo San-Jose}
\email{pablo.sanjose@csic.es}
\affiliation{Instituto de Ciencia de Materiales de Madrid (ICMM), CSIC, Madrid, Spain}
\author{Elsa Prada}
\affiliation{Instituto de Ciencia de Materiales de Madrid (ICMM), CSIC, Madrid, Spain}

\date{\today}

\begin{abstract}
We study a novel type of solitonic Andreev bound state (ABS) in a Corbino-geometry Josephson junction created on a 2DEG. The Josephson junction is subjected to a weak magnetic flux that induces a fluxoid mismatch between the inner disk and outer ring superconductors. The mismatch produces a Josephson vortex (phase soliton) that binds unconventional spinful but chargeless ABSs, analogous to Jackiw-Rebbi solitonic states. The position around the Josephson junction of the trapped ABSs can be controlled externally by a junction phase bias. As the solitonic ABSs are shuttled around the Josephson junction, the 2DEG spin-orbit coupling induces a geometric precession of their spin. We argue that these solitonic ABSs constitute a natural candidate for a novel type of superconducting Andreev spin qubit, dubbed solitonic Andreev spin qubit (SASQ), that combines features of Andreev spin qubits and geometric spin qubits. Holonomic single-qubit SASQ operations are induced through soliton shuttling, with the resulting SU(2) trajectories densely covering the qubit Bloch sphere. Effects of disorder, non-holonomic SASQ dynamics and other aspects of qubit operation are also analyzed.
\end{abstract}

\maketitle

\section{Introduction}

Qubit design plays a central role in our efforts to build a practical and universal quantum computer. The design space for qubits is large, with many factors affecting the future prospects of each qubit class.
While most leading quantum computers currently available are based on variations of the superconducting transmon qubit~\cite{Koch:PRA07,Schreier:PRB08}, it has been argued that optimally scalable quantum architectures will probably require different qubit designs with less control hardware~\cite{Dickel:18} and that allow for more efficient error correction schemes. Majorana-based topological qubits~\cite{Bravyi:AOP02,Nayak:RMP08,Sarma:NQI15} or cat-state qubits~\cite{Hann:PQ25,Putterman:N25} are prime examples of the latter. Other types of physical qubit implementations are being intensively investigated~\cite{Bruzewicz:APR19,Dutt:S07,Kok:RMP07,Casparis:NN18,Devoret:04,Seoane-Souto:24}. Some recent designs, such as the Kitaev-Sau chain~\cite{Kitaev:P01,Sau:NC12} or the Andreev spin qubit~\cite{Padurariu:PRB10,Hays:S21}, combine ingredients from several previous qubit ideas to circumvent some limitations.

\begin{figure}
   \centering
   \includegraphics[width=\columnwidth]{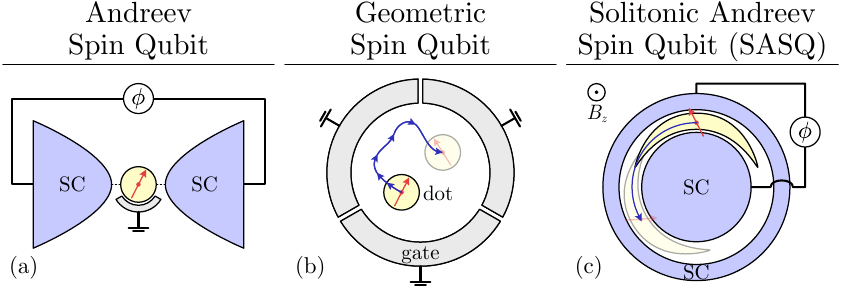}
   \caption{The solitonic Andreev spin qubit (c) combines elements from the Andreev spin qubit (a) and the geometric spin qubit (b). It is controlled with a phase bias $\phi$ like the former and allows holonomic operations like the latter. In (c), superconducting regions (blue) form a Corbino Josephson junction on a 2DEG (white). The junction is subjected to a weak out-of-plane magnetic field to induce a fluxoid mismatch between the inner disk and the outer ring. This creates a solitonic Andreev bound states (ABSs) (yellow) that are spin-degenerate, and whose position can be tuned through $\phi$. Moving the solitonic ABS induces holonomic qubit rotations due to the spin-orbit coupling in the 2DEG.}
   \label{fig:qubits}
\end{figure}

The original quantum-dot spin qubit proposal~\cite{Loss:PRA98} utilizes the spin in a quantum dot that is manipulated coherently using Zeeman fields or nanomagnets. Several qubit designs have been proposed to remove these problematic magnetic elements~\cite{Foulk:25,Heinz:25}. The Andreev spin qubit concept, in particular, replaces them with the superconducting phase difference (or a supercurrent) across a Josephson junction with the quantum dot inside; see Fig. \ref{fig:qubits}(a). Fully controlling the spin qubit this way is made possible by electrostatic gating and the presence of spin-orbit coupling in the quantum dot. Another variation of the traditional spin qubit, also reliant on spin-orbit coupling to remove the need of magnetic fields, is the geometric spin qubit concept, in which the quantum dot is moved within its two-dimensional electron gas (2DEG) host \editE{using electrostatic gates}; see Fig. \ref{fig:qubits}(b). Arbitrary manipulations of the spin in a geometric spin qubit are achieved without any magnetic fields by real-space displacements of the dot center along a path $\mathcal{C}$ in the presence of spin-orbit coupling \cite{Golovach:PRB06,Bulaev:PRL07,San-Jose:PRB08,Golovach:PRA10}.
This kind of manipulation is an example of a holonomic quantum gate~\cite{Zanardi:PLA99,Zhang:PR23,Budich:PRB12a}, which is characterized by being independent of manipulation timing (as long as it is sufficiently slow), depending purely on the geometry of $\mathcal{C}$ instead. \editE{In contrast, traditional dynamic manipulation uses electric pulses to manipulate the qubit so that its final state depends on the pulse \textit{duration}, not just the path $\mathcal{C}$ of the pulse in parameter space.} Holonomic gates remove one source of fidelity-reducing noise~\cite{Arroyo-Camejo:NC14}, and make some operations easier, such as, e.g., inverting a single-qubit gate. However, since displacements are typically small, with $\mathcal{C}$ bounded within a small region of the 2DEG, the maximum speed of holonomic operations is strongly reduced.

In this work, we study a novel type of Andreev bound state (ABS) whose properties make it a natural candidate for a new type of qubit design, dubbed solitonic Andreev spin qubit (SASQ). The SASQ combines features from the geometric spin qubit and the Andreev spin qubit; see Fig. \ref{fig:qubits}(c). In a SASQ, like in an Andreev spin qubit, single-qubit operations can be performed by controlling the superconducting phase difference across a Josephson junction. Unlike in Andreev spin qubits, however, the induced operation is holonomic. Compared to a geometric spin qubit, the possible $\mathcal{C}$ displacement paths in a SASQ can be arbitrarily large~\cite{Golovach:PRA10}, so that the holonomic qubit operations may be far more efficient, even with a weak spin-orbit coupling.

The basic elements of the SASQ are: (a) a Corbino Josephson junction, fabricated by epitaxial deposition of an inner superconductor disk and an outer superconductor ring on a 2DEG with spin-orbit coupling; (b) a weak out-of-plane magnetic field $B_z$ such that the flux threading the outer ring is around one flux quantum; and (c) a superconducting phase bias $\phi$ applied between the disk and the ring, which can be tuned either with an external \editE{superconducting quantum interference device (SQUID)} or by an electric voltage $V(t) = \dot{\phi}(t)/2e$ across the junction. \editE{As we show in this work,} these ingredients give rise to a soliton \cite{Clem:PRB10} in the $\varphi$-dependent superconducting phase difference across the junction, where $\varphi$ is the polar angle around the junction. The phase soliton (also called Josephson vortex or fluxon \cite{Barone:82,Tinkham:04}) \editE{is a localized, stable $2\pi$ twist in the superconducting phase difference. In our case, it} is the result of a mismatch between the superconductor phase winding number of the disk and the ring, caused by the magnetic flux. We find that the soliton traps a special type of Jackiw-Rebbi~\cite{Jackiw:PRD76} ABSs, concentrated around an angle $\varphi_0$, that can be controlled with $\phi$~\footnote{Majorana bound states trapped by solitons in Corbino Josephson junctions with topological superconductors have been analyzed elsewhere~\cite{Grosfeld:PNAS11,Park:PRL15}. In our case, we consider conventional (non-topological) superconductors.}. These \emph{solitonic} ABSs are spin-degenerate and implement a spin qubit when the junction is tuned to odd occupancy. The SASQ state can be holonomically controlled by tuning $\phi$, which shuttles the soliton around the junction. We show that the full Bloch sphere can be covered in this way by slightly adjusting the effective spin-orbit length $\bm{\lambda}_\text{SO}^{-1}$ through, e.g., the junction chemical potential. 


\editE{This paper is organized as follows: We describe the Corbino Josephson junction in Sec. \ref{s:model}, together with a minimal model for its electronic structure in Sec. \ref{ss:minimal}. The appearance of a soliton in the phase difference across the junction is discussed in Sec. \ref{ss:soliton}. Analytical solitonic ABS solutions are derived and analyzed in Sec. \ref{s:ABS}, with extended perturbation theory results included in Appendix \ref{ap:pert}. In Sec. \ref{s:beyond} we provide considerations beyond the minimal model, including a more general tight-binding model in Sec. \ref{ss:tb}. Effects of disorder in the junction are studied in Appendix \ref{ap:disorder}. In Sec. \ref{s:sasq} we describe how to implement a SASQ using solitonic ABSs and how to perform holonomic single-qubit operations. Non-holonomic effects are discussed in Sec. \ref{s:non-holonomic}. Finally, in Sec. \ref{s:conclusions} we conclude. A number of additional appendixes are included at the end to briefly discuss different aspects of SASQ operation, including initialization and readout (Appendix \ref{ap:ir}), dynamic (non-holonomic) single-qubit gates (Appendix \ref{ap:gates1}), two-qubit gates (Appendix \ref{ap:gates2}) and expected dephasing mechanisms (Appendix \ref{ap:decay}).}

\section{System and model}
\label{s:model}

\editE{We consider a Corbino Josephson junction consisting of a thin-film inner superconducting disk of radius $R_0$ and a narrow and thin outer superconductor ring of radius $R_1$, both s-wave and placed on top of a semiconductor quantum well hosting a 2DEG. The quantum well is shallow, so that the superconductors are able to proximitize the 2DEG immediately beneath them. A weak, millitesla scale, magnetic field $B_z$ along direction $z$ (perpendicular to the plane) is applied to the device. We assume that the separation between the disk and the ring is small compared to the superconducting coherence length, so that we can approximate $R_0\approx R_1\equiv R$. This is the so-called short-junction limit in a conventional superconductor-normal-superconductor junction. 
Given the small superconductor film thickness $d\ll \lambda_L$ (where $\lambda_L$ is the London penetration length), the screening currents that arise in response to the applied $B_z$ field are weak and thus the field can be considered unscreened. The flux threading the ring is then given simply by the external field, $\Phi = \pi R^2 B_z$, and is not quantized. As long as $R$ is smaller than the Pearl length \cite{Pearl:APL64,Clem:PRB10} $\lambda_P = 2\lambda_L^2/d$, the superconducting pairing in the disk can be assumed to be spatially uniform, $\Delta_0(\bm{r}) = \Delta_0$.  However, this is not the case for the ring's pairing $\Delta_1(\bm{r})$. Due to its doubly connected geometry, the flux $\Phi$ threading the ring induces a winding number (called \emph{fluxoid}) of the phase of $\Delta_1(\bm{r})$ as one moves around the ring. The fluxoid is quantized and is given by $n=\lfloor \Phi/\Phi_0\rceil$, where $\Phi_0=h/2e$ is the superconducting flux quantum. Fluxoid quantization was first predicted by London in multiply-connected superconductors threaded by a flux~\cite{London:50,Tinkham:04}. The constant phase in the disk and an $n\neq 0$ phase winding in the ring constitute a fluxoid mismatch.}

\subsection{Minimal model}
\label{ss:minimal}

\editE{
Our goal is to compute the Andreev spectrum of the junction. We start by presenting a minimal model for the Corbino Josephson junction that is sufficient to capture its low-energy electronic structure. It has the advantage of allowing analytical solutions for its lowest eigenstates and a clear interpretation of their topological origin. Later, in Sec. \ref{s:beyond} we provide an in-depth discussion of the physics beyond the minimal model, including a more involved numerical tight-binding model. Both models are found to be consistent, with only small deviations discussed in Sec. \ref{s:non-holonomic}.
}

\begin{figure}
   \centering
   \includegraphics[width=\columnwidth]{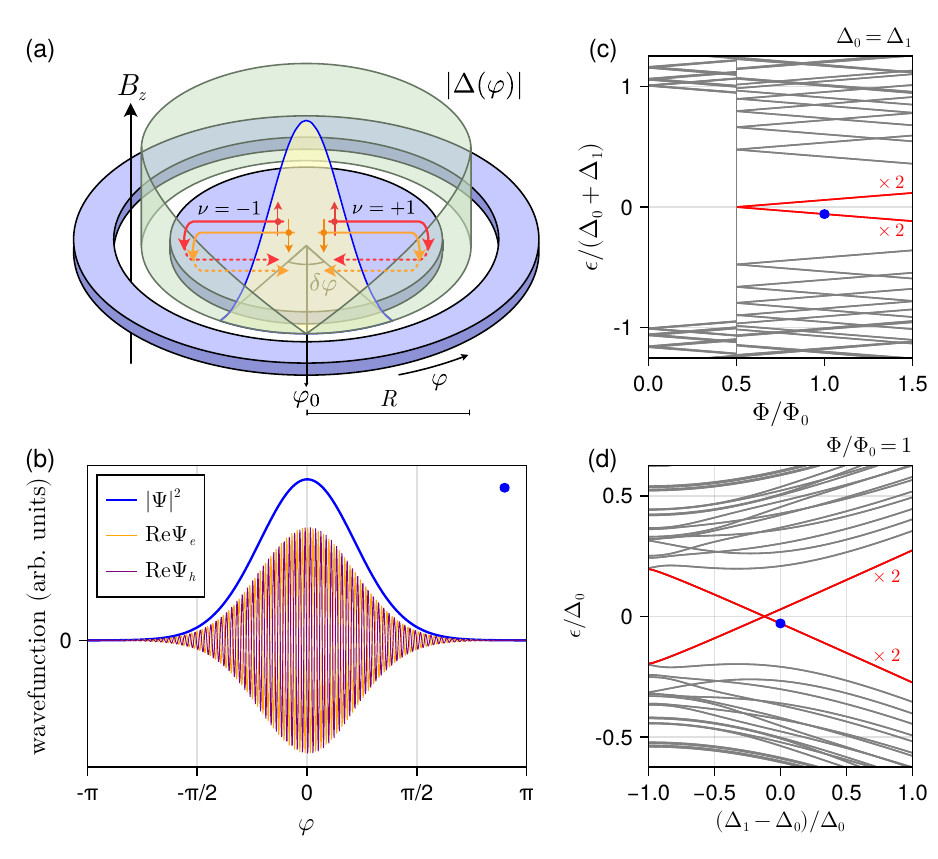}
   \caption{(a) Sketch of the solitonic ABSs (yellow) concentrated at radius $R$ within an angular spread $\delta\varphi$ from the minimum of the total pairing $|\Delta(\varphi)|$ (green) at $\varphi_0$. The $\Delta(\varphi)$ profile, Eq. \eqref{Delta}, results from a different fluxoid number in the inner disk (zero) and the outer ring (one) produced by a flux $\Phi=\pi R^2B_z$ between $0.5\Phi_0$ and $1.5\Phi_0$, where $\Phi_0$ is the superconducting flux quantum. There are two spinful low-energy states of opposite energy around each $\nu=\pm 1$ Fermi point. Their analytical wavefunction [without spin-orbit coupling, Eq. \eqref{solution}] is shown in (b) for $\Phi/\Phi_0=1$, \editE{$R = 2\mu$m} and $\Delta_0=\Delta_1=0.2$meV, where $\Delta_0$ and $\Delta_1$ are the paring amplitudes from the disk and ring. Their corresponding energy versus $\Phi/\Phi_0$ and versus pairing asymmetry $\Delta_0-\Delta_1$ (at fixed $\Delta_0+\Delta_1$) is shown in red in (c) and (d), respectively, with the rest of the spectrum in gray.}
   \label{fig:analytics}
\end{figure}

A description of the Andreev states in a Corbino Josephson junction can be obtained by integrating out the gapped superconductor regions surrounding the junction region. The latter is modeled as a one-dimensional electron gas confined in a ring of radius $R$ and polar coordinate $\varphi$; see Fig. \ref{fig:analytics}(a). We consider that the 2DEG has Fermi velocity $v_F$, Fermi momentum $k_F$ and a Rashba term $H_\text{SO} = \alpha(k_y\sigma_x - k_x\sigma_y)$, where $\alpha$ is the Rashba spin-orbit coupling, $\sigma_i$ are Pauli matrices and $\bm{k}$ is momentum (we take $\hbar=e=1$ throughout).  If we linearize the dispersion around the two Fermi momenta $\pm k_F$ along $\varphi$, we can write an effective Bogoliubov-de Gennes Hamiltonian~\cite{De-Gennes:18} around each Fermi point $\nu=\pm 1$ as
\beqa
\label{model}
H_\nu &=& \left(\begin{array}{cc}
    H_\nu^{\uparrow\uparrow} & H_\nu^{\uparrow\downarrow} \\
    H_\nu^{\downarrow\uparrow} & H_\nu^{\downarrow\downarrow}
\end{array}\right), \\
H^{\sigma\sigma}_\nu &=& \left(\begin{array}{cc}
    v_F(\nu k_\varphi - k_F - \nu A_\varphi) & \sigma \Delta(\varphi) \\
    \sigma \Delta(\varphi)^* & -v_F(\nu k_\varphi - k_F + \nu A_\varphi)
\end{array}\right),\nn\\
H^{\uparrow\downarrow}_\nu&=& (H^{\downarrow\uparrow}_\nu)^\dagger = \left(\begin{array}{cc}
    \alpha e^{i\varphi} k_\varphi & 0 \\
    0 & \alpha e^{-i\varphi} k_\varphi
\end{array}\right).\nn
\eeqa
The Nambu basis $\check{c}^\dagger = (c_\uparrow^\dagger, c_\downarrow, c_\downarrow^\dagger, c_\uparrow)$ is chosen. Here $\sigma=\pm 1$ denotes spins $\uparrow,\downarrow$, $k_\varphi=\frac{1}{R}(-i\partial_\varphi)$ denotes the momentum around the junction, and $A_\varphi=RB_z/2$ is the vector potential in the symmetric gauge along the $\varphi$ direction at the radial position $R$ for a uniform magnetic field $B_z$.  The total induced pairing $\Delta(\varphi)$ in the junction region is the sum of the disk and the ring contributions,
\beq
\label{Delta}
\Delta(\varphi)=\Delta_0 + \Delta_1(\varphi) = \Delta_0 + \Delta_1 e^{in\varphi}e^{i\phi}.
\eeq
\editE{Since $\Delta(\varphi)$ is an essential ingredient of our model, let us discuss it in more detail. As argued above, the superconducting phase of the inner disk is spatially uniform, and in particular independent of $\varphi$. This means that we assume no trapped vortices in the disk. Hence we can take $\Delta_0(\bm{r}) = \Delta_0$ as a real constant without loss of generality. The superconducting pairing around the ring is $\Delta_1(\varphi) = \Delta_1 e^{in\varphi}e^{i\phi}$ for some real $\Delta_1$ and some real $\phi$. The phase difference across the junction, $\tilde \phi(\varphi) = \arg \Delta_1(\varphi)-  \arg \Delta_0= n\varphi+\phi$, then depends on position $\varphi$. At $\varphi=0$, the phase difference is $\phi$. If we assume a phase-bias loop is attached to the ring at $\varphi=0$ [see Fig. \ref{fig:qubits}(c)], then $\phi$ can be understood as the externally controlled phase bias. If $n\neq 0$ ($|\Phi/\Phi_0|>1/2$), the magnitude $|\Delta(\varphi)|$ of the total induced pairing is $\varphi$ dependent due to the fluxoid mismatch. If $n=1$, the case we will focus on in the remainder of this work, $|\Delta(\varphi)|$ reaches a minimum $|\Delta_1-\Delta_0|$ at angle $\varphi_0 = \pi-\phi$. This is illustrated by the green curve in Fig. \ref{fig:analytics}(a). The position of the minimum is thus controlled by the phase bias $\phi$ \footnote{Closely related to this form of pairing interference is the fluxoid valve effect, recently predicted in full-shell nanowires \cite{Paya:25}. There, it was shown that a cylindrically symmetric Josephson junction will have a zero critical current when the two sides of the junction have different fluxoids. The same effect operates in our Corbino junction~\cite{Sherrill:PRB79,Clem:PRB10}.}.}

\editE{
\subsection{Phase solitons}
\label{ss:soliton}

The fluxoid mismatch between the ring and disk pairings results in a superconducting phase difference $\tilde\phi(\varphi)$ between them that makes $n$ twists around the junction. In the literature, a position-dependent superconducting phase difference along an extended Josephson junction is known as a Josephson vortex, phase soliton or fluxon. The properties and dynamics of Josephson vortices have been extensively studied \cite{Barone:82,Tinkham:04}, even in junctions with a Corbino geometry~\cite{Clem:PRB10}. Self-consistent Josephson vortices in tunnel Josephson junctions are governed by the sine-Gordon equation for $\tilde \phi$, or integral generalizations thereof~\cite{Clem:PRB10}. This equation encodes the interplay between the energy cost associated with $\tilde \phi$-phase gradients along the junction and the $\tilde \phi$-dependent Josephson energy related to the junction's critical current density $J_c$. Josephson vortices are soliton-like solutions of the self-consistent sine-Gordon equation, each of them with a winding number $\pm 1$. Their spatial extension $\lambda_J$ in sufficiently long junctions is controlled by $J_c$, as $\lambda_J\sim J_c^{-1/2}$ \cite{Tinkham:04}. In our Corbino junction $J_c$ is assumed to be sufficiently small so that $\lambda_J\gg 2\pi R$ and the $n$-soliton solution reduces to $\tilde{\phi}(\varphi) = n\varphi+\phi$, which is a self-consistent solution in this limit. In other words, in the small $J_c$ regime, the phase difference is spread uniformly around the junction. 
For $n=1$, the soliton $\tilde{\phi}(\varphi) = \varphi+\phi$ yields the $|\Delta(\varphi)|$ profile shown in Fig. \ref{fig:analytics}(a), which has a minimum at the $\pi$-junction angle $\varphi_0$ such that $\tilde\phi(\varphi_0)=\pi$.
}

\section{Solitonic Andreev bound states}
\label{s:ABS}

The model in Eq. \eqref{model} and Eq. \eqref{Delta} has two spin-degenerate ABSs $|\Psi^0_{\nu\sigma}\rangle$ per Fermi point $\nu$ close to zero energy. They are spatially localized around $\varphi_0$, as represented in yellow in Fig. \ref{fig:analytics}(a). In the special case of $\Delta_0=\Delta_1$ and $\alpha = 0$, these are also the lowest-energy eigenstates of the 2x2 $H_\nu^{\sigma\sigma}$ block, which allows us to find an exact analytical expression for their wavefunction and energy~\footnote{Note that $H_\nu^{\sigma\sigma}$ has also another eigenstate $|\bar{\Psi}_{\nu\sigma}\rangle$ with opposite energy $-\epsilon_\nu$, but this state is actually the same as $|\Psi_{-\nu\sigma}\rangle$. Thus, Eq. \eqref{solution} actually spans all four eigenstates with energy $\pm \epsilon_\nu$}:
\beqa
\langle\varphi|\Psi_{\nu\sigma}\rangle &=& C e^{-\frac{1}{2}\frac{\left(4\sin\frac{\varphi-\varphi_0}{4}\right)^2}{\delta\varphi^2}}
\left(\begin{array}{c}
    \sigma\, e^{i(\varphi-\varphi_0)(\nu k_FR+1/4)} \\ \nu\, e^{i(\varphi-\varphi_0)(\nu k_FR-1/4)}
\end{array}\right), \nn\\
\epsilon_\nu&=&-\frac{\nu v_F}{2R}\left(\frac{\Phi}{\Phi_0}-\frac{1}{2}\right).
\label{solution}
\eeqa
\editE{The above spinor lives in the $(c_\sigma^\dagger, c_{-\sigma})$ Nambu subspace.} The angular width of the states is $\delta\varphi=\sqrt{v_F/(R\Delta_1)}=\sqrt{\xi/R}$, while $\xi=v_F/\Delta$ is the superconducting coherence length and $C$ is a normalization constant. A typical wavefunction is represented in Fig. \ref{fig:analytics}(b), and the eigenenergy as a function of $\Phi$ is shown in red in Fig. \ref{fig:analytics}(c). Higher excitations, computed \editE{numerically using a full tight-binding, short-junction model (see Sec. \ref{s:non-holonomic})}, are shown in gray. The effect of a finite $\Delta_0-\Delta_1$ and $\alpha$ treated in perturbation theory is discussed in Appendix \ref{ap:pert}.

Several important aspects of the solution in Eq. \eqref{solution} stand out: (1) the eigenstates are peaked around $\varphi_0$, with an angular width given by $\delta\varphi$; (2) the states have spin but no charge, as can be seen by the equal $c^\dagger_\sigma$ and $c_{-\sigma}$ amplitude modulus of the wave function; (3) the two $\sigma = \pm 1$ excitations within each $\nu$ sector are degenerate, since $\epsilon_\nu$ is independent of $\sigma$, while states from opposite Fermi points $\nu=\pm 1$ have opposite energy; (4) the energy is also independent of $\varphi_0$ (and hence of phase bias $\phi$) and of Fermi momentum $k_F$; (5) the energy vanishes as $\Phi/\Phi_0$ approaches $0.5$ from above; (6) the low-energy Andreev levels are well separated spectrally from higher excitations, with a spacing that closely resembles the square-root dependence of Landau levels in graphene, see Fig. \ref{fig:analytics}(c) around $\Phi/\Phi_0\approx 0.5$.

The latter observation is not a coincidence. In fact, an intriguing interpretation of these states can be found by connecting the minimal model to the problem of Dirac fermions in two dimensions. By expanding around $\varphi=\varphi_0$, \editE{we get $\Delta(\varphi) \approx (\Delta_0-\Delta_1) + i(\varphi-\varphi_0)\Delta_1 + \mathcal{O}[(\varphi-\varphi_0)^2]$. We can then} map $H_\nu^{\sigma\sigma}$ to the 2D Dirac Hamiltonian (i.e., low-energy graphene~\cite{CastroNeto:RMP09}) with a position-dependent mass term \editE{(the second term in the expansion)} that changes sign at a boundary (a so-called scalar field soliton).  This is a 2D version of the celebrated 1D Jackiw-Rebbi problem~\cite{Jackiw:PRD76}. It exhibits topological fermion states confined to the boundary by virtue of the opposite valley Chern number~\cite{Thouless:PRL82,Hasan:RMP10} on each side. In the mapping to the 2D graphene version, $\sigma$ plays the role of spin, $\nu$ becomes valley, $\epsilon_\nu$ is a valley-Zeeman term, electron/hole is the pseudospin, $(\Delta_0-\Delta_1)/v_F$ is the momentum along the boundary, and $(\varphi-\varphi_0)\Delta_1$  is the position-dependent mass term, with $\varphi_0$ the position of the boundary. The mapping allows us to interpret the $|\Psi_{\nu\sigma} \rangle$ states as solitonic Jackiw-Rebbi solutions, whose existence is topologically dictated by the bulk-boundary correspondence principle, and does not require the small $\varphi-\varphi_0$ expansion used in the argument above.

\editE{
\section{Beyond the minimal model}
\label{s:beyond}

Before proceeding to the discussion of the SASQ, it is convenient to elaborate on the validity and limitations of our formalism. Our minimal model aims to capture the basic ingredients of  Corbino Josephson junctions that are essential for the emergence of solitons, solitonic Andreev states and their potential use as spin qubits. In real devices one can identify additional physical ingredients that are not captured by our model, and merit discussion. 

First, the pairings $\Delta_0$ and $\Delta_1$ were taken as constants, but should actually depend on flux $\Phi$. This dependence is a central point in the literature on the Little-Parks effect~\cite{Little:PRL62,Sabonis:PRL20}. For rings and cylinders, the dependence becomes important as soon as the radius becomes comparable to or smaller than the superconducting coherence length, which in thin aluminum films is around $\sim 100$nm, depending on the thickness and disorder. For the micron-size rings considered here, therefore, the $\Phi$ dependence of the pairing is small. In disks, the flux dependence may be more severe. 
The precise $\Delta_i(\Phi)$ dependence could be computed and incorporated into our model, but as long as $|\Delta_i|$ does not vanish, such a dependence is secondary, as it does not affect the existence and spin structure of solitonic ABSs.

Second, geometrical aspects can be relevant. We have approximated $R_0\approx R_1$. A finite junction width $R_1-R_0$ introduces an additional radial degree of freedom for junction electrons and holes. In this situation, the electrostatic potential induced by the two superconductors can vary across this junction width. The resulting Andreev spectrum can develop additional radial subbands that depend on this potential. These complications, however, only matter beyond the short-junction regime, i.e. when the junction width exceeds the coherence length (or the gap exceeds the Thouless energy). We assume the short junction limit for simplicity, since the appearance of additional radial subbands does not affect the fundamental solitonic ABSs. In this limit, the average potential induced by the superconductors is absorbed into the model's chemical potential, which defines $k_F$. 

Third, our model neglects the Josephson energy associated with the critical current density $J_c$. As explained in Sec. \ref{ss:soliton}, a finite $J_c$ can alter the profile of the $\Delta(\varphi)$ pairing, although only for large enough radius $R$. However, the specific self-consistent profile of $\Delta(\varphi)$ only affects the quantitative details of the associated solitonic ABSs. The basic qualitative aspects, including their potential use as spin qubits, depend only on the topological structure of the Josephson vortex, namely the winding $n$, not on its specific profile.

Fourth, our model neglects disorder. This is potentially an important aspect to consider, since it is ubiquitous in hybrid super-semi structures~\cite{Prada:NRP20,Ahn:PRM21} and could potentially alter the structure of solitonic ABSs, particularly their spin splittings. We devote Appendix \ref{ap:disorder} to a numerical analysis of atomic (sharp) and electrostatic (smooth) potential disorder as an extension to our results. Our main finding in this regard is that the solitonic ABSs are surprisingly insensitive to disorder, even for strong disorder comparable to the gap, particularly when it is relatively smooth, as expected from trapped charges. While the energy $\epsilon_\nu$ of the solitonic ABSs may be weakly affected by disorder, we find that it does not induce additional spin-splitting, which is the relevant aspect for spin qubit applications.

Finally, our model in Sec. \ref{ss:minimal} neglects band-curvature effects, since it relies on a linearization of the dispersion around the Fermi points. We can relax this approximation by employing instead a tight-binding description of the junction, defined below. This description is useful for example to efficiently compute higher-energy states [see gray lines in Fig. \ref{fig:analytics}(c,d)], and to quantify how dispersion curvature modifies the properties of the analytical eigensates, as discussed in Sec. \ref{s:non-holonomic}.

\subsection{Tight-binding model}
\label{ss:tb}

A tight-binding generalization of the minimal model is defined by discretizing the junction region into a large number $N$ of sites at position $\bm{r}_i$ on a circle of radius $R$, with lattice spacing $a_0=2\pi R/N$. The tight-binding Hamiltonian reads
\beq
\label{tbmodel}
H = \sum_{ij\sigma\sigma'}c_{i\sigma}^\dagger h_{ij}^{\sigma\sigma'}c_{j\sigma'} + \frac{1}{2}\left(\sum_{i\sigma}\sigma\Delta(\varphi_i)c^\dagger_{i\sigma}c^\dagger_{i,-\sigma} + \text{h.c}\right).
\eeq
The normal Hamiltonian $h_{ij}^{\sigma\sigma'}$ has onsite and nearest-neighbor terms that depend on the chemical potential $\mu$ relative to the band center, the spin-orbit coupling $\alpha$ and the hopping amplitude (or half-bandwidth) $W = 1/(m a_0^2)$, where $m$ is the effective mass of the 2DEG. These read
\beqa
h_{ii} &=& -\mu\sigma_0,\\
h_{ij} &=& e^{i \bm{A}\cdot\bm{r}_{ij}}\left(-W\sigma_0 -i\alpha\frac{\sigma_x y_{ij} - \sigma_y x_{ij}}{2a_0^2}\right),
\eeqa
where $\bm{A}$ is the vector potential between sites $i,j$, and $\bm{r}_{ij}=\bm{r}_{j}-\bm{r}_{i}=(x_{ij}, y_{ij})$ is the intersite distance vector.

The minimal model in Eq. \eqref{model} is a low-energy approximation of the above tight-binding model. The latter reduces to the former when expressed in the basis of plane waves of momentum $k_\varphi$ around the junction, by replacing the tight-binding dispersion $-W\cos\left[(k_\varphi-\nu A_\varphi)a_0\right]-\mu$ with its linearized form $v_F(\nu k_\varphi-k_F-\nu A_\varphi)$.
}

\section{Solitonic Andreev Spin Qubit}
\label{s:sasq}

\begin{figure}
   \centering
   \includegraphics[width=\columnwidth]{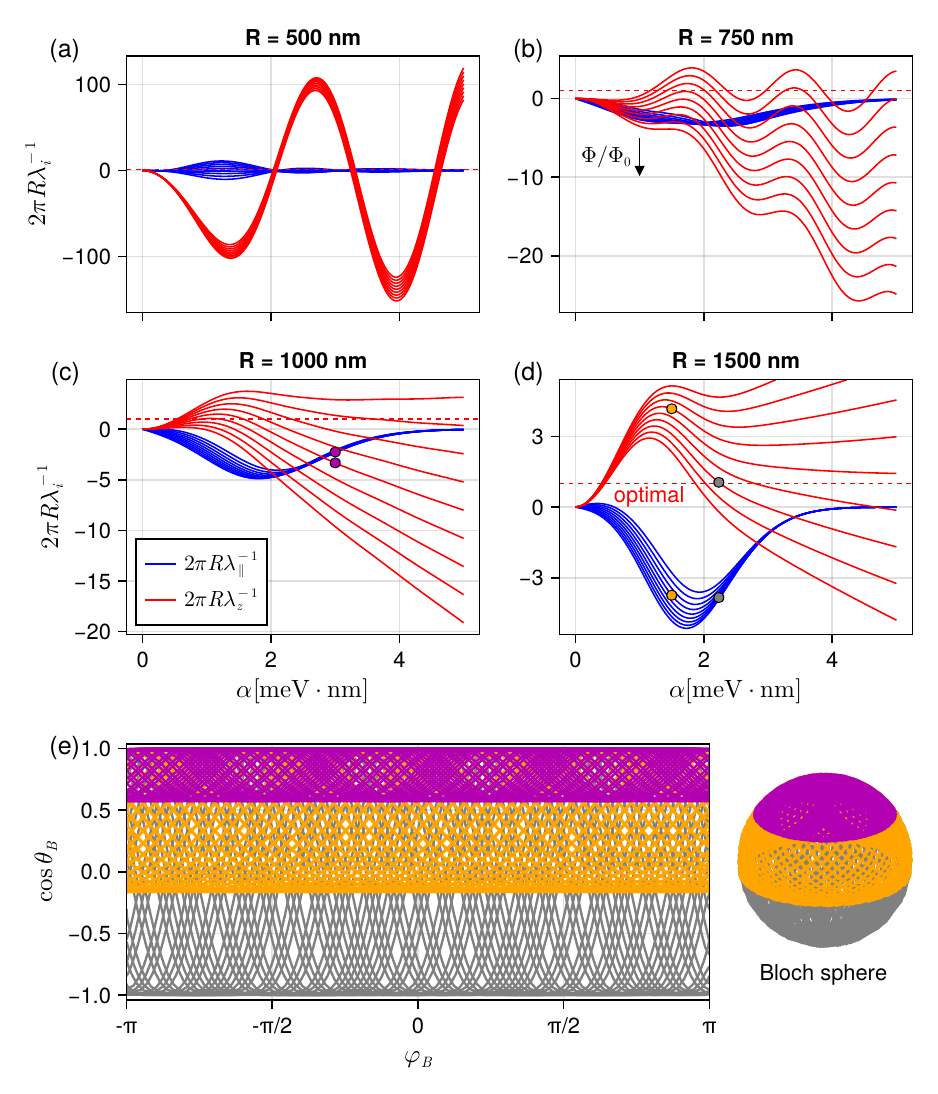}
   \caption{The two components of the inverse spin-orbit length (in-plane $\lambda_\parallel^{-1}$ in blue and out of plane $\lambda_z^{-1}$ in red), normalized to the SASQ perimeter $2\pi R$ for four different radii $R$, are shown in (a-d). Different curves correspond to growing values of normalized flux $\Phi/\Phi_0$ from 0.6 to 1.4 in steps of 0.1. The holonomic evolution of the SASQ as the solitonic ABS makes 20 turns around the junction is shown in (e) for three different values of $\alpha$ and $\Phi$ colored points in (c-d). A large fraction of the Bloch sphere, parametrized by angles $\theta_B$ and $\varphi_B$, is covered by the SASQ evolution, with full coverage (gray trajectory) achieved at the optimal point $2\pi R\lambda_z^{-1} = 1$ [dashed red line in (a-c)]. \editE{Parameters: $\Delta_0=\Delta_1=0.2$meV, $\mu = 4$meV and $N = 1000$.}}
   \label{fig:soc}
\end{figure}

\editE{We now turn to the implementation of a SASQ, a spin qubit defined by the two spin states of a solitonic ABS of fixed $\nu$. As done in quantum-dot spin qubits and Andreev spin qubits, this requires a finite charging energy in the device, which is used to fix its total fermion parity. Thus, for qubit applications, the Corbino Josephson junction should be a floating island, coupled to ground only through capacitors. Then, adjusting a capacitor voltage we can tune the Josephson junction to an odd occupancy state, which makes the ground state two-fold degenerate (the qubit), with the two associated SASQ states given by $|\Psi_{\nu\sigma} \rangle$ for fixed $\epsilon_\nu$.}

\editE{The SASQ has the remarkable feature of naturally allowing single-qubit operations using the same holonomic principle as geometric spin qubits. By tuning $\phi$ we can shift the state position $\varphi_0$ from an initial $\varphi_a$ to a final $\varphi_b$. Doing so induces a holonomic single-qubit rotation $U$ within the SASQ subspace that can be written as $U = \mathcal{P}\text{exp}\left(-\frac{i}{2}\int^{\varphi_b}_{\varphi_a} d\varphi_0 2\pi R\bm{\lambda}_\text{SO}^{-1}(\varphi_0)\cdot\bm{\sigma}\right)$~\cite{San-Jose:PRB08,San-Jose:PRL06,San-Jose:PE07}. Here,} $\mathcal{P}\text{exp}$ is the path-ordered exponential and the vector $\bm{\lambda}_\text{SO}^{-1}(\varphi_0)$ is
\beqa
\bm{\lambda}_\text{SO}^{-1}(\varphi_0) &=& \left(\lambda^{-1}_\parallel \cos\varphi_0,\lambda^{-1}_\parallel \sin\varphi_0, \lambda^{-1}_z\right) \nn \\
&=& \frac{1}{2\pi}\sum_{\sigma\sigma'}\langle\Psi_{\nu\sigma'}|k_\varphi|\Psi_{\nu\sigma}\rangle\bm{\sigma}_{\sigma\sigma'}.
\eeqa
\editE{We have parametrized $\bm{\lambda}_\text{SO}^{-1}$ in terms of in-plane and out-of-plane inverse spin-orbit length components, $\lambda^{-1}_\parallel$ and $\lambda^{-1}_z$. We are not able to derive analytical expressions for $\lambda^{-1}_\parallel$ and $\lambda^{-1}_z$ of the solitonic state with a finite $\alpha$. Instead, we compute them numerically using the full Andreev wavefunctions in the tight-binding model described in Sec. \ref{ss:tb}.} The result, normalized to the junction circumference $2\pi R$, is shown in Fig. \ref{fig:soc}(a-d) for increasing junction radius $R$. The dimensionless quantities $2\pi R\lambda^{-1}_{\parallel,z}$ are found to depend strongly on $R$, $\alpha$ and $\Phi$, reaching large values even for modest spin-orbit coupling $\alpha$ below 5~meV~nm.

The number of rotations of the SASQ after increasing $\varphi_0$ by 2$\pi$ (one turn around the disk) is of the order of $2\pi R|\bm{\lambda}_{\text{SO}}^{-1}|$, but the precise path in the Bloch sphere is complicated. Figure \ref{fig:soc}(e) \editE{shows the computed qubit paths $(\cos(\theta_B/2), \sin(\theta_B/2)e^{i\varphi_B})$ in the Bloch sphere, all starting from $| \!\!\uparrow\rangle = (1,0)$, as the solitonic ABS makes 20 full turns around the junction. Each colored path corresponds to parameters indicated by the corresponding dots in Figs. \ref{fig:soc}(c,d).} We find that by slightly tuning $\Phi$ or $\alpha$, we can realize a large variety of holonomic qubit transformations that can densely cover a large part of the Bloch sphere, in the sense that with enough revolutions we can approach any point as much as needed. Full coverage is achieved for $2\pi R\lambda_z^{-1} = 1$, see dashed red line in Fig. \ref{fig:soc}(a-d), resulting in the gray trajectory in Fig. \ref{fig:soc}(e). This kind of single-parameter holonomic manipulation \editE{with full asymptotic Bloch sphere coverage} is a peculiar possibility of the SASQ \footnote{A related result for the particular case of $\lambda^{-1}_z = 0$ (using instead arbitrarily large $R$) was obtained by Golovach \emph{et al.}~\cite{Golovach:PRA10} in the context of holonomic spin manipulations of a small quantum dot electrostatically transported around a circle.}. Alternative modes of holonomic SASQ manipulation using two or more control parameters are, of course, also possible by combining $\phi$ control with variations of other device parameters such as $k_F$, $v_F$, or $\alpha$.

\section{Non-holonomic effects}
\label{s:non-holonomic}

Up to this point, our analysis predicted spin-degenerate solitonic ABSs, which is a requirement for ideal holonomic gates on the SASQ. However, spin degeneracy is not guaranteed by any fundamental symmetry of our system (note that Kramers' theorem~\cite{Zhang:PRL13,Tasaki:20} does not apply, since time reversal symmetry is broken by the magnetic flux). Spin degeneracy in our model is a consequence of the linear dispersion. This holds even with finite spin-orbit coupling. If the dispersion acquires a finite curvature, the spin degeneracy is generally broken for finite $\alpha$, creating a finite spin splitting $\delta\epsilon^\alpha$ in the qubit subspace. A Zeeman effect from the magnetic field, albeit weak, would also contribute to $\delta\epsilon^\alpha$ (with equal or opposite sign). The spin splitting $\delta\epsilon^\alpha$ imposes a lower bound on the velocity of $\phi$-controlled qubit operations to remain holonomic. If the gate frequency is \editE{comparable to or smaller} than $\delta\epsilon^\alpha/\hbar$, the holonomic orbit will acquire a non-holonomic (dynamic) ``drift'' that depends on the driving frequency, \editE{and which should be taken into account in qubit manipulations.}

\begin{figure}
   \centering
   \includegraphics[width=\columnwidth]{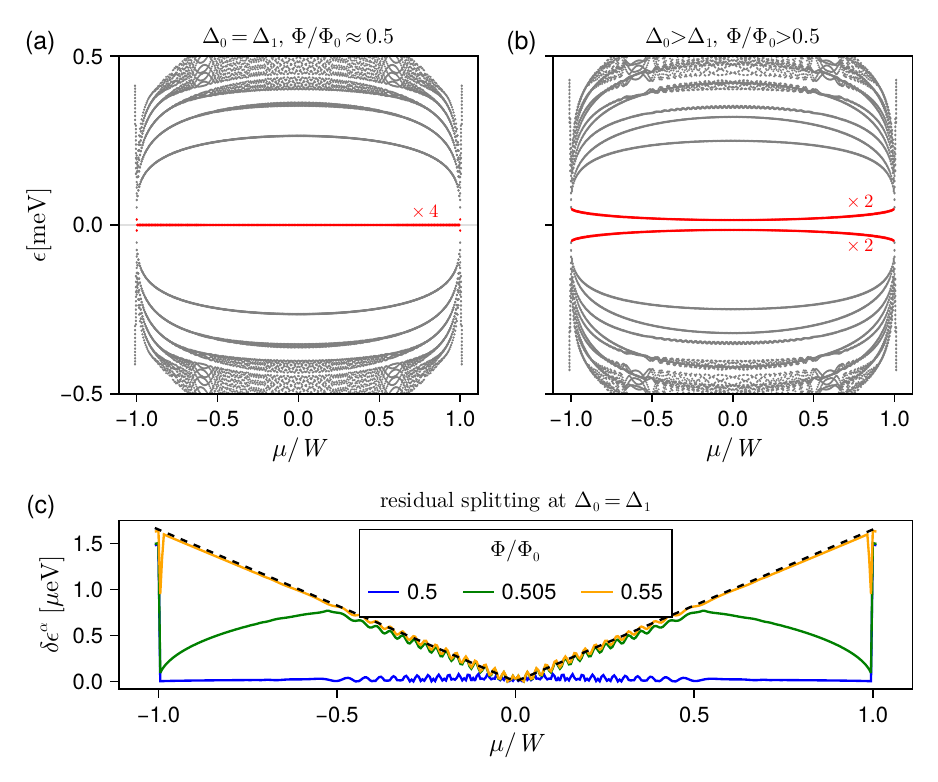}
   \caption{(a,b) Bogoliubov spectrum of a SASQ versus chemical potential $\mu$ measured from the center of a band with cosine-like dispersion, half-bandwidth $W$ and spin-orbit coupling $\alpha = 3$ meV nm. In (a), $\Delta_0=\Delta_1=0.2$ meV and $\Phi/\Phi_0\approx 0.50$. In (b) $\Delta_0=0.25$ meV, $\Delta_1=0.2$ meV and $\Phi/\Phi_0= 0.85$. Shared parameters are $W = 30$ meV, $a_0=6.3$ nm and $R = 1\mu$m.
   (c)~The non-linear dispersion and finite $\alpha$ produce a small spin splitting $\delta\epsilon^\alpha$ of the two $\pm|\epsilon_\nu|$ solitonic level pairs [red curves in (a,b)]. Colored curves are different values of $\Phi/\Phi_0$ for $\Delta_0=\Delta_1$. At half-filling $\mu=0$ and at $\Phi/\Phi_0\approx 0.5$ for any $\mu$ (blue curve), the residual splitting is suppressed. The dashed line is the analytical expression \eqref{splitting}. For $\Delta_0\neq \Delta_1$, the splitting behaves similarly to the orange curve, matching the analytic result (not shown).}
   \label{fig:mu}
\end{figure}

\editE{To compute the $\delta\epsilon^\alpha$ splitting we consider a finite curvature in the dispersion by using the tight-binding model at finite filling. Its dispersion curvature is finite throughout the bandwidth, except at its center ($\mu = 0$), where the curvature vanishes.} The Bogoliubov spectrum of the junction as a function of $\mu$ across the full band is shown in Fig. \ref{fig:mu}(a,b) for two sets of representative model parameters with a finite $\alpha = 3$meV nm. For the special case $\Delta_0=\Delta_1$ and $\Phi/\Phi_0\approx 0.5$ in panel (a), the two $\nu=\pm 1$ level pairs coalesce into a four-fold quasi-degenerate solitonic mode at zero energy across the band. In all cases, the spin splitting of the solitonic levels is small, and cannot be resolved at the meV scales shown. In Fig. \ref{fig:mu}(c) we show the splitting $\delta\epsilon^\alpha$ at zoomed-in $\mu$eV scales for several fluxes and $\Delta_0=\Delta_1$. For flux $\Phi/\Phi_0 = 0.55$ (orange line) and higher fluxes, the splitting is found to be  described rather accurately (up to a superimposed oscillation around the band center) by the following formula (dashed black line):
\beq
\label{splitting}
\delta\epsilon^\alpha \approx \frac{\alpha}{R}\frac{\Phi}{\Phi_0}\frac{|\mu|}{W}, 
\eeq
\editE{where, recall, $W$ is the half-bandwidth}. However, at smaller fluxes approaching $\Phi/\Phi_0 = 0.5$ from above (green and blue curves), the splitting is suppressed.

Two main results may be drawn. First, the residual splitting due to dispersion-curvature effects in the SASQ is upper-bounded by the scale $\alpha/R$, which is in the $\sim 1\mu$eV scale for micron-sized junctions ($\sim 1$GHz frequencies), although actual splittings can be much smaller. And second, the residual splitting and non-holonomic effects are completely suppressed when either $\Phi/\Phi_0$ approaches $0.5$ or the density of states at the Fermi energy becomes flat, as corresponds to half filling above, and to the linear dispersion in the model of Eq. \eqref{model}. \editE{In Appendix \ref{ap:disorder} we revisit these results in the presence of non-magnetic disorder.}

\section{Conclusions}
\label{s:conclusions}


\editE{
We have analyzed Andreev states bound to a soliton in a Corbino-geometry Josephson junction on a 2DEG and their potential for a SASQ design. A magnetic flux applied to the junction makes the superconducting phase wind differently as one moves around the outer ring than in the inner disk. This difference, known as a fluxoid mismatch, gives rise to a soliton in the phase difference across the junction. We showed that a phase soliton traps pairs of ABSs, which are spin-degenerate up to corrections from the dispersion curvature. Spin-orbit coupling endows them with a spin texture that leads to a spin precession as they are shuttled around the junction by moving the soliton with a phase bias. This constitutes the basis for the SASQ qubit concept and its holonomic manipulation. The optimal radius $R$ for the Corbino junction is $\xi \lesssim R \ll \lambda_P$, so that it is large enough to remain in the non-destructive Little-Parks regime and to trap a flux quantum with a weak magnetic field, but it is still small enough so that no Pearl vortex, of size $\lambda_P$, can enter the inner disk. This corresponds to radii in the $\sim 0.5-2\mu$m range for aluminum.

Solitonic ABSs are a very special kind of Andreev state. We clarified their topological origin by connecting them to the Jackiw-Rebbi problem. Experimentally, one could aim to detect them in a Corbino junction and demonstrate their solitonic character by Andreev spectroscopy. The Andreev spectrum with a fluxoid mismatch is profoundly different than without. Two near-zero excitations, spectrally isolated from the rest of the spectrum, should emerge as we induce a fluxoid mismatch in the junction (see Fig. \ref{fig:analytics}). Detecting the emergence of such ABSs should constitute strong evidence of their solitonic nature. Conclusive evidence could be obtained through local spectroscopy (using, e.g., a quantum point contact close to a point in the outer ring perimeter) by proving that the ABS states are localized and that a phase or voltage bias across the junction leads to their shuttling around the junction.

The simplicity of the proposed device, the lenient requirements regarding magnetic fields and spin-orbit coupling, the expected insensitivity to electrostatic disorder, the possibilities afforded by the real-space shuttling of the soliton, and the maturity of available techniques to grow shallow, epitaxially proximitized quantum wells, make solitonic ABSs and SASQ devices an appealing target of future experiments.}

\acknowledgments{
We are grateful to Fernando Barbero, Charles M. Marcus, Jelena Klinovaja and Daniel Loss for useful discussions.
This research was supported by Grants PID2021-122769NB-I00, PID2021-125343NB-I00 and PID2024-161665NB-I00 funded by MICIU/AEI/10.13039/501100011033, ``ERDF A way of making Europe'' and ``ESF+''. We acknowledge the Severo Ochoa Centres of Excellence program through Grant CEX2024-001445-S, and the CSIC’s Quantum Technologies Platform (QTEP).
}

\textit{Data availability---} The code required to generate the data and the plots  in this study is available through Zenodo at Ref. \onlinecite{San-Jose:Z26}, and is built using the Quantica.jl package \cite{Quantica}.

\appendix

\section{Perturbation theory}
\label{ap:pert}

The effect of a finite $\Delta_0-\Delta_1$ on the exact $\alpha=0$ solution of Eq. \eqref{solution} can be computed in perturbation theory. This pairing asymmetry does not break the spin degeneracy, and simply amounts to an energy shift, with no change to eigenstates,
\beqa
\delta\epsilon^\Delta_\nu &=& (\Delta_0-\Delta_1)\nu\gamma, \\
\gamma &=& \frac{\int_{-\pi}^\pi d\varphi e^{-\left(4\sin\frac{\varphi}{4}\right)^2/\delta\varphi^2} \cos\frac{\varphi}{2}}{\int_{-\pi}^\pi d\varphi e^{-\left(4\sin\frac{\varphi}{4}\right)^2/\delta\varphi^2}} = \nn \\
&=& \frac{I_1(8/\delta\phi^2)+L_{-1}(8/\delta\phi^2)}{I_0(8/\delta\phi^2)+L_{0}(8/\delta\phi^2)},
\eeqa
where $I_n$ and $L_n$ are the $n$-th order modified Bessel function of the first kind and the modified Struve function. The expression for $\gamma$ reduces to $1$ and $2/\pi$ in the limits of small and large $\delta\varphi$, respectively. The eigenenergies, computed numerically as a function of $\Delta_1-\Delta_0$, are shown in red in Fig. \ref{fig:analytics}(d), which show that perturbation theory is accurate. A similar perturbative calculation for the spin-orbit coupling yields zero change of the eigenenergies to first order in $\alpha$ (although the eigenstates do change). Spin degeneracy is thus preserved at this level.

\begin{figure}
   \centering
   \includegraphics[width=\columnwidth]{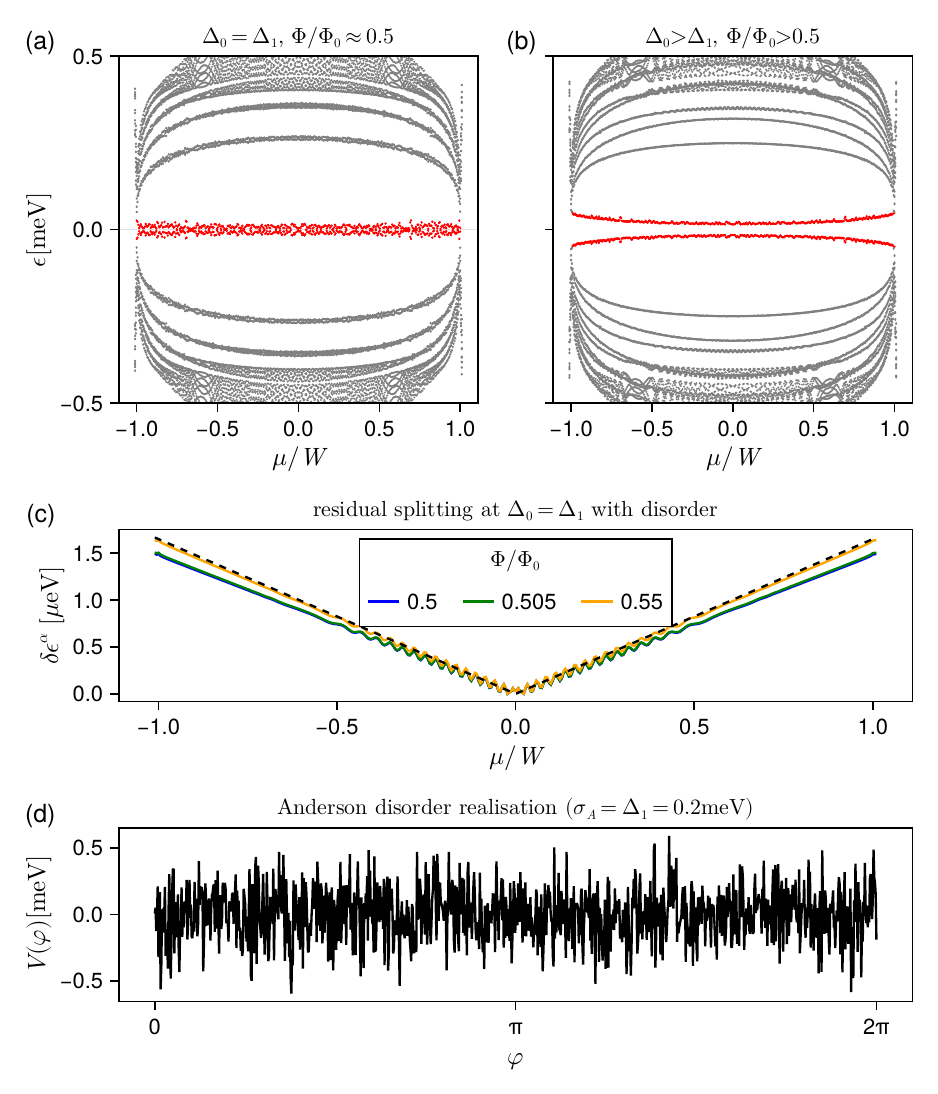}
   \caption{\editE{Bogoliubov spectrum and spin splitting of the solitonic ABSs analogous to Fig. \ref{fig:mu}, but in the presence of a single realization of Anderson disorder, shown in (d), with standard deviation $\sigma_A=0.2$meV.}}
   \label{fig:Anderson}
\end{figure}

\begin{figure}
   \centering
   \includegraphics[width=\columnwidth]{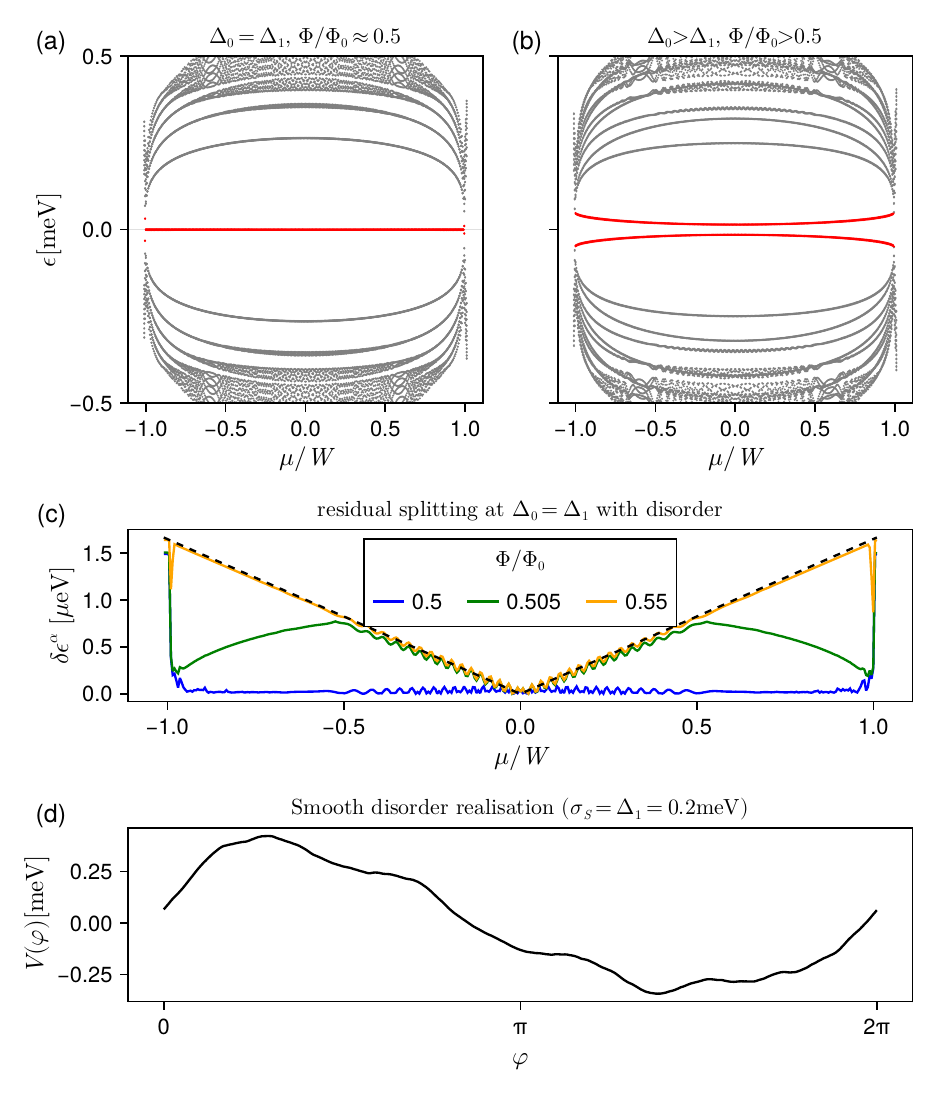}
   \caption{\editE{Bogoliubov spectrum and spin splitting of the solitonic ABSs analogous to Fig. \ref{fig:mu}, but in the presence of a single realization of smooth disorder, shown in (d), with standard deviation $\sigma_S=0.2$meV.}}
   \label{fig:smooth}
\end{figure}

\editE{
\section{Effects of non-magnetic disorder}
\label{ap:disorder}

Hybrid devices that combine conventional superconductors and semiconductors have been often found to suffer from charge disorder~\cite{Prada:NRP20,Ahn:PRM21}, particularly due to trapped charges in the dielectric environment, which is difficult to control. In this Appendix we revisit the tight-binding calculations of Sec. \ref{s:non-holonomic} including a random, position-dependent, non-magnetic potential in the junction to assess the sensitivity to disorder of the SASQ energy and spin-splitting.

In our tight-binding Hamiltonian, an additional non-magnetic disorder term takes the form
\beq
H_D = \sum_i^N\sum_\sigma V(\varphi_i) c^\dagger_{i\sigma}c_{i\sigma},
\eeq
where $N$ is the total number of sites, typically $N=1000$ in our simulations. We consider Anderson disorder and smooth disorder models for $V(\varphi)$. In the Anderson model, on the one hand, the random potential on each site $i$ is $V(\varphi_i) = W_i$,
where $W_i$ are independent, Gaussian, zero-average, random variables with standard deviation $\sigma_A$, $\overline{W_iW_j} = \delta_{ij}\sigma_A^2$. Conceptually, this disorder model represents uncorrelated, atomic-like dislocations and impurities inside the junction. On the other hand, the smooth disorder model represents trapped charges in the dielectric environment, which induce a smoothed-out random potential landscape on the junction electrons. We model this landscape as $V(\varphi) = \sqrt{2}\mathrm{Re}\sum_{k=1}^N V_k e^{i k \varphi}$, where the $V_k$ harmonics are complex, Gaussian random variables with a variance that decays strongly with harmonic order $k$, $\overline{V_k^*V_{k'}} = \delta_{kk'}\sigma_S^2/k^2$. The local variance of $V(\varphi)$ can be shown to be $\overline{V(\varphi)^2} = (\pi^4/90)\sigma_S^2 \approx \sigma_S^2$, and its average is also zero.

In Figs. \ref{fig:Anderson} and \ref{fig:smooth} we show the Bogoliubov spectrum and spin splitting in the presence of Anderson and smooth disorder, respectively. We consider strong disorder in both cases comparable to the gap, with $\sigma_A = \sigma_S = 0.2$meV. The figure structure is analogous to that of the clean case in Fig. \ref{fig:mu}. We find that Anderson disorder results in comparatively small fluctuations (much smaller than $\sigma_A$, arguably due to self-averaging) of the total solitonic ABS energy as a function of chemical potential $\mu$. Unlike in the clean case, the spin splitting is no longer zero at $\Phi/\Phi_0\to 0.5$ and $\Delta_0=\Delta_1$. Instead, $\delta\epsilon^\alpha$ saturates to the upper bound of Eq. \eqref{splitting} for any flux. However, the spin splitting is still perfectly suppressed at the band center where the dispersion curvature vanishes. This underlines the result that the spin splitting of solitonic ABSs is a direct consequence of dispersion curvature, even in the presence of strong disorder and spin-orbit coupling.

In the more relevant case of smooth charge disorder, we find exactly the same phenomenology for the energy and spin splitting as in the clean case. The corresponding low-energy spectrum and splittings, shown in Fig. \ref{fig:smooth}, are virtually indistinguishable from Fig. \ref{fig:mu}. We see that in this case, the spin-splitting at $\Phi/\Phi_0 \to 0.5$ remains suppressed even in the presence of large potential variations with $\sigma_S=0.2$meV.

}

\section{Initialization and readout}
\label{ap:ir}

The initialization and readout of a SASQ can be done similarly to other types of spin qubits. The first step is to fix the SASQ ground state to odd occupancy. This typically requires a sufficient charging energy, \editE{which results from electrically decoupling the device from ground (a so-called floating island setup). Its occupation and parity can then be controlled with a gate capacitor}. Once loaded with a single quasiparticle, the solitonic state can be initialized to a fixed spin in two standard ways. One approach is to allow the SASQ to relax under a strong Zeeman field. Another strategy is to use an ancilla spin qubit, initially in a singlet state, to provide the quasiparticle. The ancilla will then become entangled with the initial SASQ state, and can later be used for readout. The readout is performed through spin-to-charge conversion using, e.g., spin-Coulomb blockade~\cite{Weber:NN14,Yoneda:NC20}. Attempting to transfer the SASQ quasiparticle back into the ancilla will only be possible if the two quasiparticles have a finite singlet amplitude. The degree of charge transfer can be detected, e.g., with a dispersive readout of a coupled microwave resonator frequency, which is highly sensitive to minute charge differences~\cite{DAnjou:PRB19}. \editE{These techniques are well known in the field of spin qubits, and are not specific to the SASQ architecture.}

\section{Non-holonomic single-qubit gates}
\label{ap:gates1}

We have shown that general single-qubit gates can be implemented in a SASQ through holonomic manipulation of the soliton position. The maximum angular velocity $\partial_t\varphi_0$ of the soliton should be limited by the closest excitation in the same $\nu$ sector, which according to Fig. \ref{fig:analytics}(c) is approximately $(\Delta_0+\Delta_1)/2$, or around 20ps per revolution (300GHz) for $\sim 0.2$meV gaps.

Alternatively, however, one can also implement dynamic (non-holonomic) single-qubit gates via electron-dipole spin resonance (EDSR)~\cite{Golovach:PRB06}, as in conventional spin qubits. To this end, one must induce a spin splitting $\delta\epsilon$ exceeding temperature along a given spin direction $\bm{n}$. This can stem from a combination of the dispersion-curvature effect $\delta\epsilon^\alpha$
and an external Zeeman field $V_Z = \frac{1}{2}g\mu_B |\bm{B}|$, where $\bm{n}$ is then fixed by $\bm{B}$. A small, resonantly oscillating voltage bias $V_b(t) = V_0\sin(\delta\epsilon^\alpha t)=-\partial_t\varphi_0(t)/2e$ across the Corbino Josephson junction will induce a small-amplitude motion $\varphi_0(t)$ of the soliton position around the initial $\varphi_0(0)$, which in turn induces an effective oscillating Zeeman term in the effective Hamiltonian of the form $(\partial_tU)U^\dagger = -\frac{1}{2}2\pi R\partial_t\varphi_0(t)\bm{\lambda}_\text{SO}^{-1}\bm{\sigma}$. The spin component orthogonal to $\bm{n}$ will induce Rabi oscillations. Assuming $\bm{B}=B\hat{x}$  (in-plane), the Rabi precession frequency in the rotating wave approximation reads $\omega_R = e V_0 2\pi R \sqrt{\lambda_\parallel^{-2}\sin^2\varphi_0+\lambda_z^{-2}}$. With a resonant drive $e V_0 = \delta \epsilon^\alpha\approx 20\mu$eV with $R\approx\lambda_z/2\pi \approx \lambda_\parallel/2\pi \approx 1\mu$m, we expect a maximum Rabi frequency of around $40$GHz. We therefore expect holonomic single-qubit operations to be potentially faster than non-holonomic ones.

\section{Two-qubit gates}
\label{ap:gates2}

\editE{Two-qubit gates for a pair of SASQs side by side} can be implemented following the original Loss-DiVincenzo approach~\cite{Loss:PRA98}, which relies on two nearby qubits with an odd parity, fixed by charging energy, that interact via exchange through a tunable barrier. The exchange $J\bm{S}_1\cdot\bm{S_2}$ between the SASQ spins $\bm{S}_i$, acting during a specific pulse time $\sim 1/J$, can implement a $\sqrt{\text{SWAP}}$ or a CPHASE gate~\cite{Loss:PRA98}. Combining such operations appropriately with single-qubit Hadamard and Pauli-Z gates results in a CNOT gate~\cite{Barenco:PRA95}, which is commonly chosen as part of a universal set of qubit gates.

Two-qubit gates between a SASQ interacting with an adjacent SASQ to its right can be implemented by parking the soliton in the latter at the point $\varphi_0=\pi$ closest to the first SASQ. Then, performing a sweep of the former across $\varphi_0=0$ of maximum exchange coupling $J$ at a specific speed implements a $\sqrt{\text{SWAP}}$ or CPHASE gate~\cite{Zou:PRR23}. The remaining single-qubit gates required to perform a CNOT operation could be achieved by decoupling the two SASQs, and making a specific number of soliton revolutions in the target SASQ. The latter should be tuned beforehand (through the flux, radius and/or Rashba) so that some integer number $n_Z$ and $n_H$ of holonomic revolutions correspond to the required Pauli-Z and Hadamard gates, respectively.

\section{Dephasing}
\label{ap:decay}

One of the key advantages of holonomic qubit operations is that by suppressing the qubit splitting, spin-orbit coupling-mediated spin dephasing from electric noise is typically also suppressed, since contributions to the dephasing rate $T_2^{-1}$ linear in the coupling to the environment cancel out. This is known as van Vleck cancellation~\cite{Van-Vleck:PR40}. Second-order geometric contributions then dominate~\cite{San-Jose:PE07, San-Jose:PRB08}. These too, however, should also be strongly suppressed in a SASQ, since the corresponding dephasing rate is a function of the area covered by the random electrically-induced motion of the qubit wavefunction. In a narrow Corbino Josephson junction, this area is negligible, as voltage fluctuations produce only one-dimensional random motion. We therefore expect the dominant spin dephasing mechanism in a SASQ to be exchange with magnetic impurities or the hyperfine coupling to spinful nuclei, averaged over the (large) extent of the solitonic ABS wavefunction.

\bibliography{biblio}

\end{document}